\documentclass{Interspeech2024}
\usepackage{multirow}
\usepackage{authblk}
\usepackage[utf8]{inputenc}

\interspeechcameraready

\title{Enhancing Modal Fusion by Alignment and Label Matching for Multimodal Emotion Recognition}

\name[]{Qifei}{Li}
\name[]{Yingming}{Gao}
\name[]{Yuhua}{Wen}
\name[]{Cong}{Wang}
\name[]{Ya}{Li}

\address{
  School of Artificial Intelligence, Beijing University of Posts and Telecommunications, China}
\email{\{liqifei, yingming.gao, 2020212213, congwang, yli01\}@bupt.edu.cn}


\begin{document}

\maketitle

\begin{abstract}
    
    To address the limitation in multimodal emotion recognition (MER) performance arising from inter-modal information fusion, we propose a novel MER framework based on multitask learning where \textbf{f}usion \textbf{o}ccurs \textbf{a}fter a\textbf{l}ignment, called Foal-Net. The framework is designed to enhance the effectiveness of modality fusion and includes two auxiliary tasks: audio-video emotion alignment (AVEL) and cross-modal emotion label matching (MEM). First, AVEL achieves alignment of emotional information in audio-video representations through contrastive learning. Then, a modal fusion network integrates the aligned features. Meanwhile, MEM assesses whether the emotions of the current sample pair are the same, providing assistance for modal information fusion and guiding the model to focus more on emotional information. The experimental results conducted on IEMOCAP corpus show that Foal-Net outperforms the state-of-the-art methods and emotion alignment is necessary before modal fusion. The code is open-source\footnote{https://github.com/ASolitaryMan/Foal-Net}.
\end{abstract}

\section{Introduction}

Emotion recognition is an important part of human-computer interaction (HCI). In order to improve the interaction experience, it is necessary to fully utilize the information between different modalities to improve the recognition performance of the system~\cite{tellamekala2023cold}. Hence, efficient modal fusion methods is one of the current research hotspots.

In the MER task, there are many challenges that need to be investigated. This challenges include the need for improved feature representation, developing reasonable model structure to fit the feature representation, exploring effective modal fusion methods to realize inter-modal information complementary and improve the performance of MER, and addressing or alleviating the problem of model performance degradation caused by missing modal information.


The commonly used modalities for emotion recognition include audio, video, and text. In recent years, most of researchers have extracted deep representations of pre-trained models to represent modal information~\cite{tavernor23_interspeech, tran23c_interspeech, zhao23b_interspeech, ghosh23b_interspeech}, such as HuBERT~\cite{hsu2021hubert}, WavLM~\cite{chen2022wavlm}, BERT~\cite{kenton2019bert}, RoBERT~\cite{liu2019roberta}, Resnet-FER2013~\cite{zhong2020facial} and MAE~\cite{he2022masked}. Compared with the conventional features, such as Mel Frequency Cepstrum Coefficient, one-hot representation and Face Action Unit, the deep representations have better performance and generalization. For modeling the modal information, Wang et al.~\cite{wang2023exploring} proposed a modality-sensitive MER framework to exploring complementary features. Mitra et al.~\cite{mitra2023pre} utilized TC-GRU which consists of time convolution network and GRU to capture the spatio-temporal properties of the inputs. Maji et al.~\cite{maji2023multimodal} proposed a cross-model Transformer model to model the information of audio and text modalities and achieved remarkable performance. The cross-attention mechanism is currently the most commonly used for modal fusion~\cite{vaswani2017attention}. The above studies all made use of cross-attention to fuse the audio and lexical information to realize MER. Lian et al.~\cite{lian2023gcnet} exploited graph neural networks to fuse three modalities and attained excellent performance. In order to alleviate the effects of modal absence, Lian et al.\cite{lian2023gcnet} and Wang et al.~\cite{wang2023exploring} employed different strategies individually to simulate scenarios of modality absence during the training process, enhancing the robustness of the emotion recognition model.

Recently, remarkable performance have been achieved in research related to utilizing auxiliary tasks for modality fusion. In many image-text tasks, such as image-text retrieval, visual entailment and visual question answering (VAQ), many researchers always take advantage of auxiliary tasks to assist image-text information to fuse~\cite{li2021align, radford2021learning, li2022blip, wang2023image, bao2022vlmo}. These auxiliary tasks include image-text matching, image-text contrastive learning, maskd language model, etc. In particular, the authors of ALBEF~\cite{li2021align} point out that it is very effective and necessary to perform modal alignment prior to modal fusion. Similarly, in the MER task, Ghosh et al.~\cite{ghosh23b_interspeech} and Sun et al.~\cite{sun2023using} have respectively proposed different auxiliary tasks for multimodal emotion recognition, both achieving outstanding performance.
\begin{figure*}[ht]
    \centering
    \includegraphics[scale=0.75]{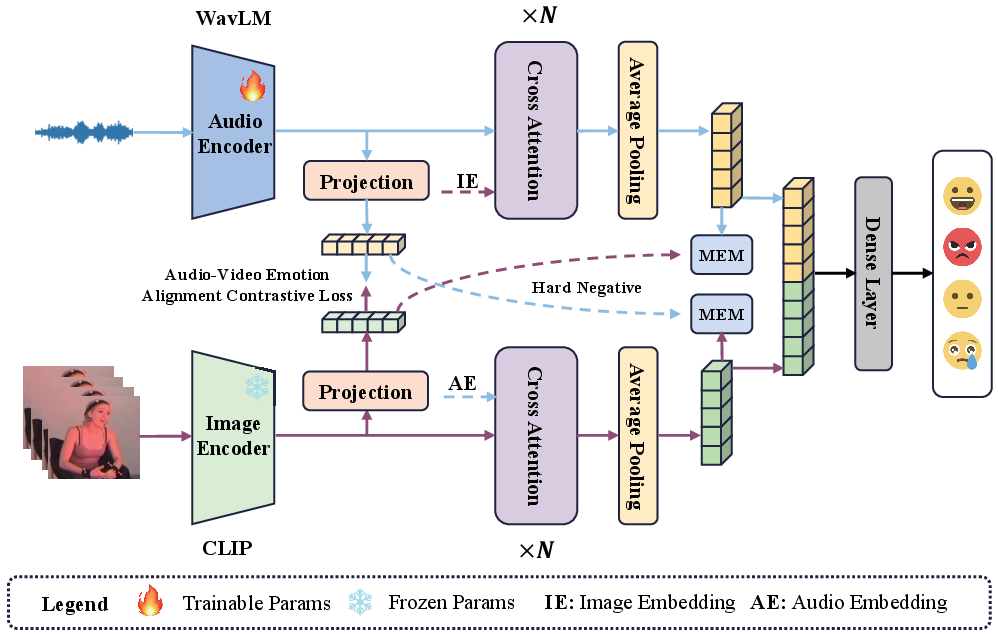}
    \caption{An overview of Foal-Net architecture, which mainly consists of two auxiliary tasks (AVEL, MEM), and modal fusion module.}
    \label{fig:fig1}
\end{figure*}

Inspired by the above studies, we propose a novel MER framework based on multitask learning for MER, named Foal-Net, which consists of two unimoal sub-networks, multiple layers of cross-attention and two auxiliary tasks. The outputs of two sub-networks first need to undergo audio-visual emotion alignment, where the audio-visual emotion alignment contrastive loss (AVEL) is computed. Subsequently, they are fed into the cross-attention network for modalities fusion. Simultaneously, we calculate negative samples for each current modality sample based on AVEL's similarity matrix and hard negative technique, forming pairs of positive-negative samples. Next, these sample pairs, along with the initially matched pairs, are used to implement corss-modal emotion label matching (MEM). The MEM serves two purposes. Firstly, it promotes modality fusion, preventing insufficient integration of modal information and recognizing the emotion of the current sample solely based on a single modality. Secondly, it directs the model to focus more on the emotional information within the current input. For modal information representation, we use CLIP embeddings instead of traditional facial expression features. This is because these embeddings not only encompass facial information but also include body language information. We have validated the effectiveness of our proposed method on the IEMOCAP corpus. Experimental results indicate that the performance of Foal-Net surpasses other SOTA methods. Our main contributions can be summarized as follows:

\begin{itemize}
\item Unlike other existing studies that mainly focus on modal fusion methods, we propose a method that involves alignment before fusion for multimodal emotion recognition, which fills the gap between current researches.
\item We introduce an emotion label matching auxiliary task to assist modal information fusion. In addition, it can pay more attention to emotional information in the fusion process.
\item The unweighted accuracy (UA) and weighted accuracy (WA) of Foal-Net are 80.1\% and 79.45\%, outperforming the other state-of-the-art (SOTA) methods.
 \end{itemize}

\section{Proposed Method}
In this section, we will provide a detailed introduction to the proposed AVEL, MEM auxiliary tasks, and the modal fusion module used.

\subsection{Audio-Video Emotion Aligning}
Inspired by ALBEF~\cite{li2021align}, in order to achieve better modal fusion for complementary emotional information between modalities in the later stage, we first align the emotional information between the audio and video modalities. In other words, the goal of AVEL is to enhance the similarity between sample pairs with the same emotional category through contrastive learning, while reducing the similarity between sample pairs with different emotional categories.

The input audio-video sample pairs are $\left\{X^a_i, X^v_i\right\}$, where $i \in{\left[0,N\right]}$, $N$ is batch size and $a, v$ represent audio, video respectively. First, Foal-Net leverages WavLM and CLIP to extract audio embeddings $Z^a \in{\mathbb{R}^{N\times T \times D_a}}$ and video embeddings $Z^v \in{\mathbb{R}^{N\times F \times D_v}}$, where $T, F$ represent the number of frames in the audio and video, respectively, $D_a, D_v$ are the dimension of audio and video embeddings. Then, we conduct average pooling along with time dimension of $Z^a$ and $Z^v$ to obtain $\bar{Z}^a \in{\mathbb{R}^{N\times D_a}}$ and $\bar{Z}^v \in{\mathbb{R}^{N\times D_v}}$. The $\bar{Z}^a$ and $\bar{Z}^v$ are fed into projection blocks, which are designed to map their inputs to feature vectors of the same dimension, enabling the calculation of inter-modal similarity matrices. These operations are measured as:
\begin{figure}[t]
  \centering
  \includegraphics[scale=0.8]{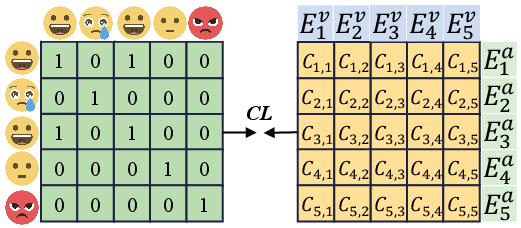}
  \caption{The example for AVEL. The horizontal and vertical axis represents the video and audio modality respectively. As indicated in the subplot on the left, the value is 1 when their emotional labels are same; otherwise, it is 0. $CL$ denotes contrastive loss.}
  \label{fig:fig2}
\end{figure}
\begin{align}
  E^a = MLP_a{(\bar{Z}^a)}; E^v = MLP_v{(\bar{Z}^v)}
  \label{equation:eq1}
\end{align}
\begin{align}
  C^{a2v} = \epsilon \times (E^a \cdot E^{v^{T}}); C^{v2a} = \epsilon \times (E^v\cdot E^{a^{T}})
  \label{equation:eq2}
\end{align}
where the $MLP_a(\cdot)$ and $MLP_v(\cdot)$ are projection modules, which consist of two linear layers. $E^a\in{\mathbb{R}^{N\times D}}$ and $E^v\in{\mathbb{R}^{N\times D}}$ denote the scaled audio and video feature vectors, the $\epsilon$ is temperature hyper-parameter. The $C^{a2v}\in{R}^{N\times N}$ and $C^{v2a}\in{R}^{N\times N}$ represent inter-modal similarity matrices.

The labels $\tilde{C}^{e}\in{R}^{N\times N}$for the contrastive loss, as shown in Figure 2, are set such that if the true labels are the same for different sample pairs within the same batch, the corresponding ground truth values are set to 1. Otherwise, these ground truth values are set to 0. Finally, the inter-modal emotion alignment loss is defined as:
\begin{align}
  L_a &= -\frac{1}{2}(\frac{\sum^{N}_{i=1}\sum^{N}_{j=1}(\sigma(C^{a2v}_{i,j})\tilde{C}^{e}_{i,j} + \sigma(C^{v2a}_{i,j})\tilde{C}^{e}_{i,j})}{N})
  \label{equation:eq3}
\end{align}
where the $\sigma$ is log-softmax function.
\subsection{Audio-Video Emotion Matching}
The research by Sun et al.~\cite{sun2023using} indicates that during the training process, for some simple samples, the model can identify emotions using only the information from a single modality. This may result in insufficient fusion of modal information. Therefore, errors are more likely to occur when the model predicts complex samples. 

Hence, we propose the MEM auxiliary task to alleviate this issue. MEM is a binary classification task designed to enable the model to fully leverage information from both modalities to determine whether the emotional labels for the current input sample pair are consistent. We adopt the hard negative technique to  generate the negative samples required for binary classification. Hard negative refers to identifying samples in other modalities that have inconsistent emotional information with the current modality sample but the highest similarity. First, we will set the value of $C^{a2v}$ and $C^{v2a}$ to negative infinity at the corresponding position where $\tilde{C}^{e}$ has a value of 1. Then, the steps for finding the most similar negative sample from other modalities for the current sample are as follows:
\begin{align}
  id^{a2v}_{i} = argmax(softmax(C^{a2v}_{i})); i\in{\left[0, N\right]}
  \label{equation:eq4}
\end{align}
\begin{align}
  id^{v2a}_{i} = argmax(softmax(C^{v2a}_{i})); i\in{\left[0, N\right]}
  \label{equation:eq5}
\end{align}
\begin{align}
  Z^a_{neg} = Z^a\left[id^{v2a}\right]; Z^v_{neg} = Z^v\left[id^{a2v}\right]
  \label{equation:eq6}
\end{align}
where $id^{a2v}$ and $Z^v_{neg}$ respectively represent the index and CLIP embeddings of samples in the video modality. They are selected to serve as negative samples for the audio modality. Similarly, $id^{v2a}$ and $Z^a_{neg}$ represent the index and WavLM embedding of samples from the audio modality, serving as negative samples for the video modality. Then, they are fed into fusion network for MEM. The operations are as follows:
\begin{align}
  M^a_p=\delta(f_a(Z^a, Z^v)); M^a_n=\delta(f_a(Z^a, Z^v_{neg}))
  \label{equation:eq7}
\end{align}
\begin{align}
  M^v_p=\delta(f_v(Z^v, Z^a)); M^v_n=\delta(f_v(Z^v, Z^a_{neg}))
  \label{equation:eq8}
\end{align}
\begin{align}
  L_m = \frac{1}{2}\sum^{k\in{\left\{a,v\right\}}}_{k}CE(cat(M^{k}_{p}, M^{k}_{n}), \tilde{M}_e)
  \label{equation:eq9}
\end{align}
where $f_a(\cdot)$ and $f_v(\cdot)$ denote fusion networks for audio and video modalities respectively. The $\delta$ means average pooling along with time dimension. The $CE$ represents \textit{Cross Entropy} loss. $M^k_p$ and $M^k_n$ ($k \in{\left\{a,v\right\}}$) denote the output of paired positive and unpaired negative samples from the fusion network, respectively. $\tilde{M}_e \in {\mathbb{R}^{2N\times 1}}$ represents the labels used for MEM, where the first half takes the value of 1, and the second half takes the value of 0.

Finally, the loss function of Foal-Net is represented as Equation~\ref{equation:eq10}, where $L_{ce}$ is the emotion classification loss, and the value of $\lambda$ is 0.01.
\begin{align}
  L_{Total} = L_{ce} + L_a + \lambda L_m
  \label{equation:eq10}
\end{align}

\subsection{Modal Fusion Module}
The fusion module we used is based on multi-head cross-attention~\cite{vaswani2017attention}. Here, we use the representation of the speech modality as the query to illustrate the calculation method of the cross-attention mechanism. The calculation process of multi-head cross-attention is as follows:
\begin{align}
  F^{a}_{out} = LN(MH(Q^{a}, K^{v}, V^{v})+F^{a})
  \label{equation:eq11}
\end{align}
\begin{align}
  MH(Q^{a}, K^{v}, V^{v}) = C({head}_1,\cdots,{head}_h)
  \label{equation:eq12}
\end{align}

\begin{align}
  {head}_h &= Attention(Q^{a}_h, K^{v}_h, V^{v}_h) \nonumber \\
       &= softmax(\frac{Q^{a}_h(K^{v}_h)^T}{\sqrt{d_k}})(V^{v}_h)
\end{align}

\begin{align}
  Q^{a}_h = F^{a}W^Q_h + b^Q_{h}
\end{align}

\begin{align}
  K^{v}_h = F^{v}W^K_h + b^K_{h}
\end{align}

\begin{align}
  V^{v}_h = F^{v}W^V_h + b^V_{h}
\end{align}
where $F^{a}$ and $F^{v}$denotes the embeddings from pretrained audio and video models respectively. $LN$, $MH$ and $C$ mean Layer Normalization, Multi-Head and Concatenate. The $h$ represents the $hth$ cross-attention. When calculating $F^v_{out}$, the method remains consistent with $F^a_{out}$, but the query is now based on the representation of the video modality. In this paper, we utilize two multi-head cross-attention layers for modal fusion, with each layer consisting of four heads. The $F^a_{out}$ and $F^v_{out}$ will undergo an average pooling layer along with time dimension, and then be concatenated together for MER.

\section{Experiments and Results}
\subsection{Experimental Setup}
\textbf{Dataset} The IEMOCAP~\cite{busso2008iemocap} multimodal corpus is one of the most well-known databases for emotion recognition, encompassing data from three modalities: audio, video, and text. In total, it comprises 5 sessions, each including one male and one female speaker. To be consistent and compare with previous studies, we conduct experiments with 5,531 audio utterances of four emotion categories \textit{happy} (\textit{happy \& excited}, 1,636), \textit{angry} (1,103), \textit{sad} (1,084) and \textit{neutral} (1,708). We perform five-fold cross-validation using a leave-one-session-out strategy on the corpus to evaluate the effectiveness of our proposed method. The weighted accuracy (WA) and unweighted accuracy (UA) are used as metrics in line with previous methods.

\textbf{Experimental Details} The feature dimensions of the WavLM\footnote{https://huggingface.co/microsoft/wavlm-large} and CLIP\footnote{https://huggingface.co/openai/clip-vit-large-patch14} image encoders are 1024 and 768, respectively. The Projection module has 512 neurons, with a dropout rate of 0.5. During training, the batch size is set to 64, the learning rate is a constant 1e-4, and the optimizer is AdamW. The dropout rate for cross-attention is 0.1. The input for the audio modality consists of 6 seconds of speech with a sampling rate of 16 kHz. Due to the simultaneous appearance of two speakers in the video recording, it is necessary to separate the speakers in the frame. In the end, the input for the video modality consists of 180 individual images. Any questions regarding the parameters in the method we proposed, please refer to the github repository. 

\begin{table}[t]
  \centering
  \caption{Performance comparison between the proposed model and other SOTA methods on IEMOCAP corpus. The A means Audio, V denotes Video.}
  \label{tab:table_1}
    \begin{tabular}{@{}ccccc@{}}
    \toprule
    Methods  & Type & Year & UA(\%) & WA(\%) \\ \midrule
    FTSLLM~\cite{yang2022exploiting}        & A+T  & 2023 & 78.50      & 77.70      \\
    MSMSER~\cite{wang2023exploring}        & A+T  & 2023 & 76.40      & 75.20      \\
    ATMF~\cite{sun2023using}        & A+T  & 2023 & 79.71      & 78.42      \\
    RMER~\cite{lin2023robust}        & A+T  & 2023 & 77.00      & 76.00      \\ 
    MTTX~\cite{He10097110}        & A+T  & 2023 & 75.00      & 74.50      \\
    CRNN-SAN~\cite{maji2023multimodal}        & A+T  & 2023 & 79.95      & 78.82      \\
    BAM~\cite{zhao2023knowledge}        & A+T  & 2023 & 77.00      & 75.50      \\
    MSER~\cite{khan2024mser}        & A+T  & 2024 & 76.56      & 77.20      \\ \midrule
    MMAN~\cite{pan20b_interspeech}        & A+V  & 2020 & -      & 73.94      \\
    AM-FBP~\cite{zhou2021information}        & A+V  & 2021 & 75.49      & -      \\
    MCWSA-CMHA~\cite{zheng2022multi}        & A+V  & 2022 & 78.90      & -      \\ 
    GCNet~\cite{lian2023gcnet}        & A+V  & 2023 & 78.36      & -      \\ \midrule
    \textbf{Foal-Net} & A+V  & -    & \textbf{80.10}       & \textbf{79.45}       \\ \bottomrule
    \end{tabular}
\end{table}

\subsection{Performance Comparison with Previous Methods}
The effectiveness of the method we proposed can be underscored by contrasting it with the latest significant findings derived from the IEMOCAP corpus, as show in Table~\ref{tab:table_1}. It demonstrates that the best UA (80.10\%) and WA (79.45\%) are achieved by Foal-Net. Research on facial expression recognition in the IEMOCAP corpus is relatively scarce, primarily due to two reasons. Firstly, the video frames are relatively blurry, making it challenging to extract facial expression features. Secondly, the issue of speakers appearing in the same frame in the videos complicates the data processing. However, based on the experimental results, the performance of the audio-video combination is not inferior to that of the audio-text combination. This suggests that besides facial information, other details in the images also contribute to emotion recognition. Meanwhile, it highlights the outstanding performance of the Foal-Net.

\subsection{Experiments and Analysis}
We conducted a series of ablation experiments to validate the effectiveness of the features we used, the proposed auxiliary tasks, and the final model. The baseline is the Foal-Net without two auxiliary tasks. 
To validate the effectiveness of global image encoding for emotion recognition compared to local facial information encoding, we extracted deep representations from three pretrained models (EmoNet~\cite{kahou2016emonets}, Resnet-FER2013~\cite{zhong2020facial}, SENet-FER2013~\cite{zhong2020facial}) for facial expression recognition and utilized them for emotion recognition. As shown in Table~\ref{tab:table_2}, their performance is significantly lower than that of the CLIP model's features. This indicates that, in addition to facial expressions, body movements and other information in the images also contribute to emotion recognition. Furthermore, another reason for the superior performance of CLIP features is the inclusion of semantic information.

\begin{table}[t]
  \centering
  \caption{The results of ablation experiments related to feature input, auxiliary tasks, and modal fusion conducted on the IEMOCAP corpus.}
  \label{tab:table_2}
    \begin{tabular}{@{}cccc@{}}
    \toprule
    Type                 & Methods              & UA(\%)               & WA(\%)               \\ \midrule
    \multirow{6}{*}{V}   & \textbf{Face-only}                  & \multicolumn{1}{l}{}                    & \multicolumn{1}{l}{}                   \\
                         & EmoNet                  & 37.22                    & 36.50                    \\
                         & SENet-FER2013                  & 36.18                    & 35.53                    \\
                         & Resnet-FER2013                 & 40.74                    & 40.16                    \\
                         & \textbf{All Image}                    & \multicolumn{1}{l}{}                    & \multicolumn{1}{l}{}                   \\
                         & CLIP                  & 47.90                    & 46.61                    \\ \midrule
    A                    & WavLM & 77.68 & 77.25 \\ \midrule
    \multirow{4}{*}{A+V} & Baseline                  & 77.85                    & 77.31                    \\
                         & Baseline+AVEL                  & 79.31                    & 78.20                    \\
                         & Baseline+MEM                  & 78.10                    & 76.94                    \\
                         & \textbf{Baseline+AVEL+MEM}                  & \textbf{80.10}                    & \textbf{79.45 }                   \\ \bottomrule
    \end{tabular}
\end{table}

In addition, we observe that after incorporating the AVEL task, UA and WA improved by 2.2\% and 1.36\%, respectively, compared to the baseline. This indicates the crucial necessity of cross-modal emotion alignment before modal fusion. On the contrary, the performance improvement is not very significant when introducing the MEM task alone. This is because the use of hard negative for finding negative samples heavily relies on the similarity matrix from the AVEL task. When the similarity matrix is not optimized, it cannot bring positive benefits to the MEM task. This also indirectly confirms the necessity of alignment before fusion. When we introduce both the AVEL and MEM tasks simultaneously, the model's performance is further improved compared to using AVEL or MEM alone. This demonstrates that with the assistance of AVEL, MEM can facilitate the fusion of modal information and enhance the model's capability for emotion recognition.

\section{Conclusions}
In this paper, we propose a novel framework called Foal-Net for MER, which includes two auxiliary tasks: AVEL and MEM. Our research on the AVEL task has demonstrated the effectiveness and necessity of initially aligning emotional information across modalities before integrating inter-modality information. The MEM task can guide modality fusion and make fusion module focus more on emotional information during the fusion process with the assistance of the AVEL task. Under the influence of both tasks, Foal-Net achieves SOTA performance. Moreover, we show that the embeddings of CLIP outperform facial expression features in MER. In the future work, we will optimize the MEM task as its performance relies on AVEL and validate the universality of our method by substituting the video modality with the text modality.

\section{Acknowledgements}

\ifinterspeechfinal
     The work was
\else
     The work was
\fi
supported by the National Natural Science Foundation of China (No.\ 62271083), the Key Project of the National Language Commission (No.\ ZDI145-81), the Fundamental Research Funds for the Central Universities (No.\ 2023RC13, No.\ 2023RC73), BUPT Excellent Ph.D. Students Foundation (No.\ 2023116), and partly supported by the Major Program of the National Social Science Fund of China (13\&ZD189).

\bibliographystyle{IEEEtran}
\bibliography{mybib}

\end{document}